\documentclass[a4paper, aps, preprint]{revtex4}
\usepackage[latin2]{inputenc}
\usepackage[dvips]{graphicx}
\usepackage[british]{babel}

\newcommand{\Int}[2]
{\int_{#1}^{#2}\!}
\begin{document}
\title{Marked signal improvement by stochastic resonance for aperiodic signals in the double-well system}

\author{R\'obert Mingesz}
\author{Zolt\'an Gingl}
\author{P\'eter Makra}
\affiliation{Department of Experimental Physics, University of Szeged, D\'om t\'er 9, Szeged, H-6720, Hungary}

\begin{abstract}
On the basis of our mixed-signal simulations we report significant stochastic resonance induced input-output signal improvement in the double-well system for aperiodic input types. We used a pulse train with randomised pulse locations and a band-limited noise with low cut-off frequency as input signals, and applied a cross-spectral measure to quantify their noise content. We also supplemented our examinations with simulations in the Schmitt trigger to show that the signal improvement we obtained is not a result of a potential filtering effect due to the limited response time of the double-well dynamics.
\end{abstract}


\maketitle

\section{Introduction}
\label{sec:Introduction}

Having originated in the context of ice ages \cite{Benzi81}, stochastic resonance (SR) is nowadays often given a signal processing interpretation: noise aids a weak signal to surpass some kind of barrier in a system, which is then reflected in the noise content of the output of the system. As the most widely used measure of this noise content is the signal-to-noise ratio (SNR), one quantitative definition of SR may be a noise-induced optimisation of the output SNR. Stochastic resonance in itself means using noise to make the output less noisy than it would be without noise, yet the signal processing approach just mentioned impels one to to raise the question whether, in the framework of SR, noise can also make the output less noisy \emph{as compared to the input}, similarly to the way filters do. This question has long intrigued researchers working in the field of SR, and after a few unsuccessful attempts at the beginning, SR-induced input-output improvement has been demonstrated in a wide range of systems from a simple level crossing detector \cite{Kiss96, Loerincz96} and other static non-linearities \cite{ChapeauBlondeau97} to the Schmitt trigger \cite{Gingl00} and even dynamical systems such as neuronal models \cite{Liu01} or the archetypal double-well potential \cite{Gingl01}.

To our present knowledge it seems unlikely that stochastic resonance will ever rival filters in technical applications designed to improve signal quality. Yet there may exist processes, like neural signal transfer, where SR represents the only viable method of amplifying sub-threshold stimuli, and a number of findings do point in this direction \cite{Russell99, Hidaka00}. As aperiodic signals are native in this class of processes, studying their role in SR-induced signal improvement is not at all without relevance.

Quantifying the noise content of aperiodic signals poses a special problem, as the most widely used definition of the signal-to-noise ratio is valid in the strict sense only for harmonic signals, and even its wide-band extension depends on the condition of periodicity in the input signal. Several cross-correlational and cross-spectral measures have been in use in the field of aperiodic SR to circumvent this problem; here we adopt the cross-spectral treatment used by L B Kish \cite{Kiss96}.

In our present study, we apply a mixed-signal simulation environment to examine whether aperiodic signals---a randomised pulse train and a band-limited noise as signal---can also be improved by SR occurring in the archetypal double-well system.

\section{Modelling and methods}
\label{sec:ModellingAndMethods}

\subsection{Measures of noise content}
\label{sec:MeasuresOfNoiseContent}

The first to introduce the technical notion of signal-to-noise ratio into SR research were Fauve and Heslot when reporting stochastic resonance in a real-world electronic device, the Schmitt trigger \cite{Fauve83}. SNR then became widely adopted as the quantifier of SR, most often taken in the following form to facilitate analytical treatment \cite{McNamara89}:
\begin{equation}\label{eq:SNRDef}
R := \frac{\lim_{\Delta f \to 0}\Int{f_0 - \Delta f}{f_0 + \Delta f}S(f)df}{S_N(f_0)},
\end{equation}
wherein $f_0$ is the frequency of the signal, $S(f)$ denotes the power spectral density (PSD) of the signal and $S_N(f)$ stands for the background noise PSD. This definition solely concerns the immediate neighbourhood of the first spectral peak, thus, strictly speaking, it yields an appropriate description of noise content only in the case of sinusoidal signals. In our papers, we have argued for the adoption of a more practical SNR interpretation favoured in electronics, which takes into account all spectral peaks and the whole background noise power:
\begin{equation}\label{eq:SNRwDef}
R_w := \frac{\sum_{k = 1}^\infty \lim_{\Delta f \to 0}\Int{k f_0 - \Delta f}{k f_0 + \Delta f}S(f)df}{\Int{0}{\infty}S_N\left(f\right)df}.
\end{equation}
This definition (to which, contrasting it to the narrow-band scope of the definition in Eq \ref{eq:SNRDef}, we refer as the \emph{wide-band SNR}) is valid for non-sinusoidal periodic signals as well, and, as we have demonstrated in \cite{Gingl01}, it provides a much more realistic account of signal improvement even in the case of a sine input.

For all measures of noise content, the chief difficulty lies in separating signal from noise. Sometimes, especially in the case of the narrow-band definition in Eq (\ref{eq:SNRDef}), this was carried out by recording the PSD of the output when noise was fed into the input without any signal. This method doubled the simulation workload, as each simulation step was to be repeated without input signal, while its validity was also questionable from a theoretical point of view, since it failed to take into account the cross-modulation between signal and noise which occurs due to the non-linearity inherent in systems showing SR. In most cases, when the background noise PSD is smooth, signal-noise separation may be simplified by taking the noisy spectra as a whole and calculating the PSD of the noise at the signal frequency (or the integer multiples of the signal frequency in the case of the wide-band definition) as the average of PSD values in the neighbourhood of the spectral peak (excluding the peak itself, of course); the signal PSD is then the PSD at the spectral peak minus this averaged noise background.


When the input signal is aperiodic, neither of the above-mentioned methods works, because signal power is not concentrated at particular frequencies. This case calls for a more elaborate technique of signal-noise separation, which is usually based on either the cross-correlation (as, for example, in \cite{Collins95}) or the cross-spectrum between the noiseless input and the noisy signal. Here we reach back to the treatment used in \cite{Kiss96}, and take the signal PSD at the output as the part of the total PSD which shows correlation with the noiseless input, reflected in their cross-spectrum:
\begin{equation}\label{eq:SOut}
S_{out}^{sig}\left(f\right) = \frac{\left|S_{in,\ out}\left(f\right)\right|^2}{S_{in}^{sig}\left(f\right)},
\end{equation}
where $S_{in,\ out}\left(f\right)$ denotes the cross power spectral density of the input signal and the \emph{total} output, while $S_{in}^{sig}\left(f\right)$ is the PSD of the input signal. As the input signal and the noise are uncorrelated, the noise component of the output can be obtained simply as
\begin{equation}\label{eq:SOutNoise}
S_{out}^{noi}\left(f\right) = S_{out}^{tot}\left(f\right) - S_{out}^{sig}\left(f\right),
\end{equation}
where $S_{out}^{tot}\left(f\right)$ is the PSD of the total output. The cross-spectral SNR at the output is then defined as
\begin{equation}\label{eq:SNRcsOut}
R_{cs,\, out} := \frac{\Int{0}{\infty}S^{sig}_{out}\left(f\right)df}{\Int{0}{\infty}S^{noi}_{out}\left(f\right)df}.
\end{equation}
As we are interested in input-output signal improvement, we also need a cross-spectral SNR at the input:
\begin{equation}\label{eq:SNRcsin}
R_{cs,\, in} := \frac{\Int{0}{\infty}S^{sig}_{in}\left(f\right)df}{\Int{0}{\infty}S^{noi}_{in}\left(f\right)df},
\end{equation}
wherein
\begin{equation}\label{eq:SInNoise}
S_{in}^{noi}\left(f\right) = S_{in}^{tot}\left(f\right) - S_{in}^{sig}\left(f\right),
\end{equation}
and $S_{in}^{tot}\left(f\right)$ denotes the PSD of the \emph{total} input.

The measures we have chosen to reflect signal improvement are the \emph{signal-to noise ratio gains}, defined simply as the ratios of the output and input values of the two kinds of SNR we consider:
\begin{equation}\label{eq:WideGain}
G_w := \frac{R_{w,\, out}}{R_{w,\, in}},
\end{equation}
and
\begin{equation}\label{eq:CSGain}
G_{cs} := \frac{R_{cs,\, out}}{R_{cs,\, in}}.
\end{equation}

\subsection{The mixed-signal simulation environment}
\label{sec:TheMixedSignalSimulationEnvironment}

We modelled the archetypal dynamical SR system in which the overdamped motion of a particle in a double-well potential is given by the following Langevin equation:
\begin{equation}\label{eq:DW}
\frac{dx}{dt} = x\left(t\right) - x^3\left(t\right) + p\left(t\right) + w\left(t\right),
\end{equation}
wherein $p\left(t\right)$ denotes the input signal and $w\left(t\right)$ stands for the noise (a physical white noise---that is, having a limited bandwidth---in our case). Comparing the noise content of the input and output signals, we looked for a signal improvement induced by stochastic resonance.

\begin{figure}[htbp]
\centering
\includegraphics[width = 0.8 \textwidth]{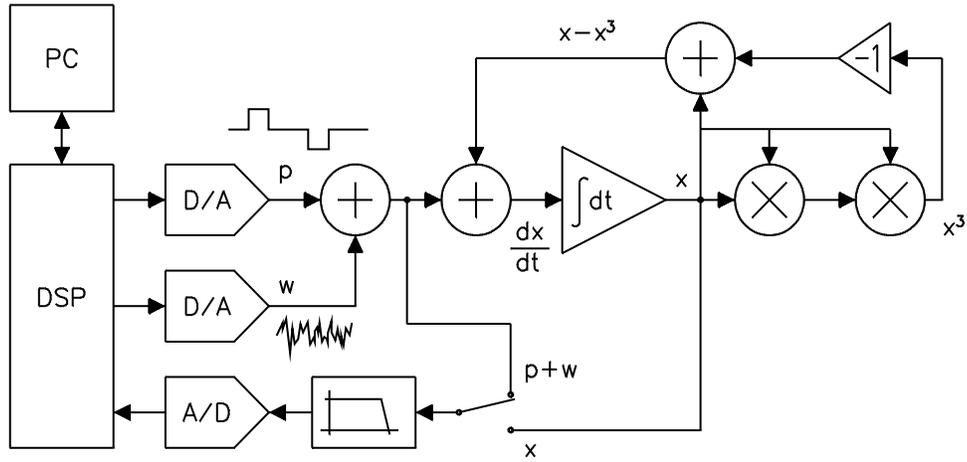}
\caption{Our mixed-signal system realising the double-well model}
\label{fig:SimulationSystem}
\end{figure}

We applied a mixed-signal (ie, having both digital and analogue components) simulation system to realise the double-well potential and solve Eq~(\ref{eq:DW}). To obtain the solution of the latter, we first transformed it into an integral form:
\begin{equation}\label{eq:DWInt}
x\left(t\right) = \Int{0}{t}\left\{x\left(\tau\right) - x^3\left(\tau\right) + p\left(\tau\right) + w\left(\tau\right)\right\}d\tau.
\end{equation}
We generated the input signal and the noise digitally, then converted them into analogue signals. All mathematical operations in Eq~(\ref{eq:DWInt}), such as addition, multiplication and integration, were performed by analogue devices. The output of our analogue circuitry was the solution of Eq~(\ref{eq:DWInt}), which we then transmitted through an anti-aliasing filter and converted back to the digital domain using high-resolution A/D converters. In order to avoid artefacts that might stem from different treatment, we used the very same filtering and sampling unit to digitise both the input and the output. The simulation system was driven by a high-performance digital signal processor (DSP), under the control of a computer running LabVIEW, which also performed all evaluation tasks. Our mixed-signal simulation system is summed up in Fig \ref{fig:SimulationSystem}.

It is worth noting that the analogue integrator introduces a $1/(RC)$ factor into Eq~(\ref{eq:DWInt}), wherein $R$ and $C$ are the resistance and the capacitance in the integrator circuit. The output of the integrator is therefore not exactly the solution $x(t)$ but
\begin{equation}\label{eq:IntegratorOutput}
y(t) = \frac{1}{b} \Int{0}{t}\left[y(\tau) - y^3(\tau) + p(\tau) + w(\tau)\right]d\tau,
\end{equation}
wherein $b := RC/(\mathrm{1\ s})$. Substituting $s := \tau \cdot 1/b$, we see that the integrator transforms the time scale by a $b$ factor:
\begin{equation}\label{eq:realdw}
y(t) = \Int{0}{t/b}\left[y(bs) - y^3(bs) + p(bs) + w(bs)\right]d s.
\end{equation}
This means that the actual frequency scale in the analogue circuitry is $1/b$ times the theoretical frequency scale corresponding to Eqs (\ref{eq:DW}) and (\ref{eq:DWInt}). In our simulations, the value of $b$ was $1.2 \cdot 10^{-4}$.

\begin{figure}[!htb]
\centering
\includegraphics[width = 0.3\textwidth]{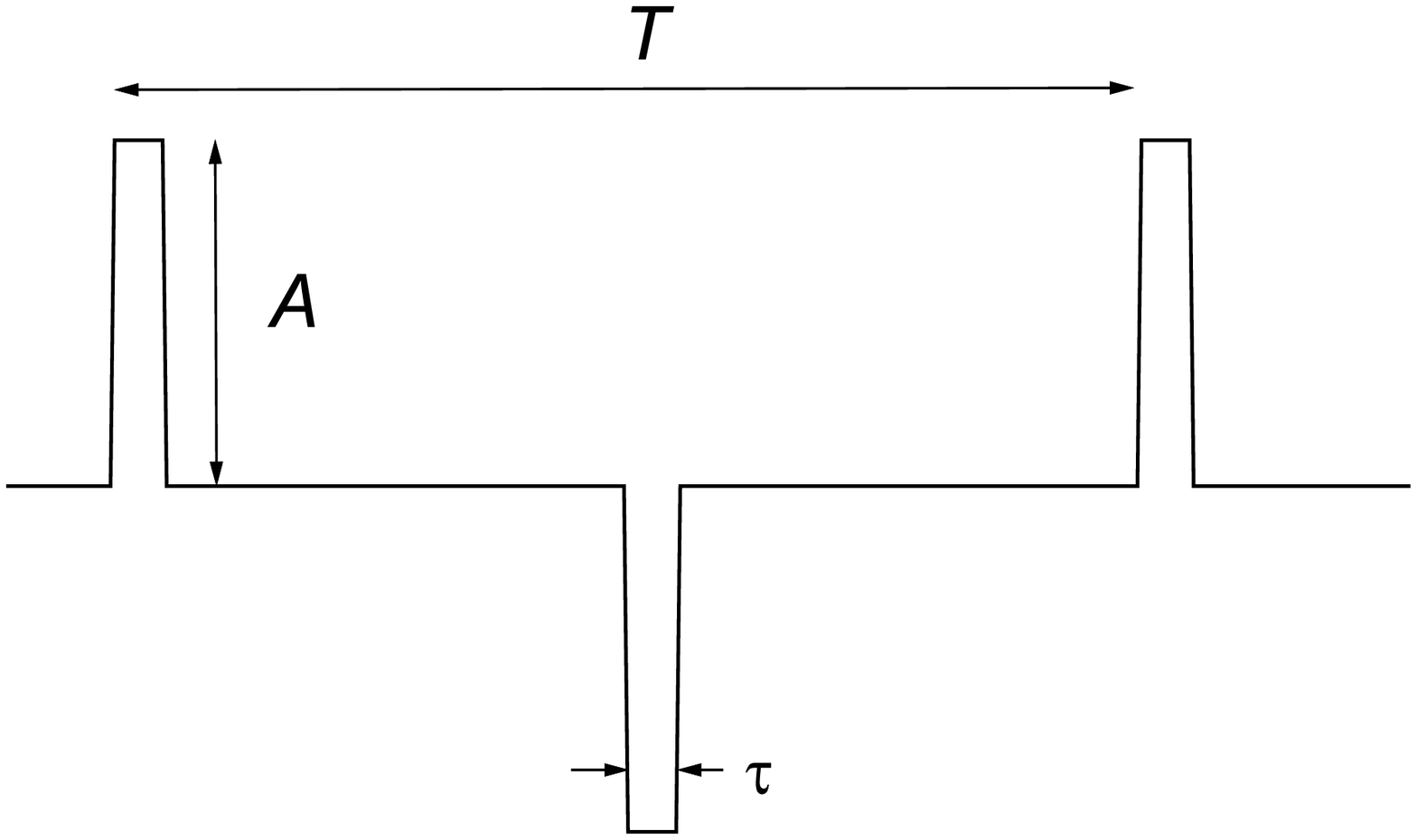}
\hspace{1 mm}
\includegraphics[width = 0.3\textwidth]{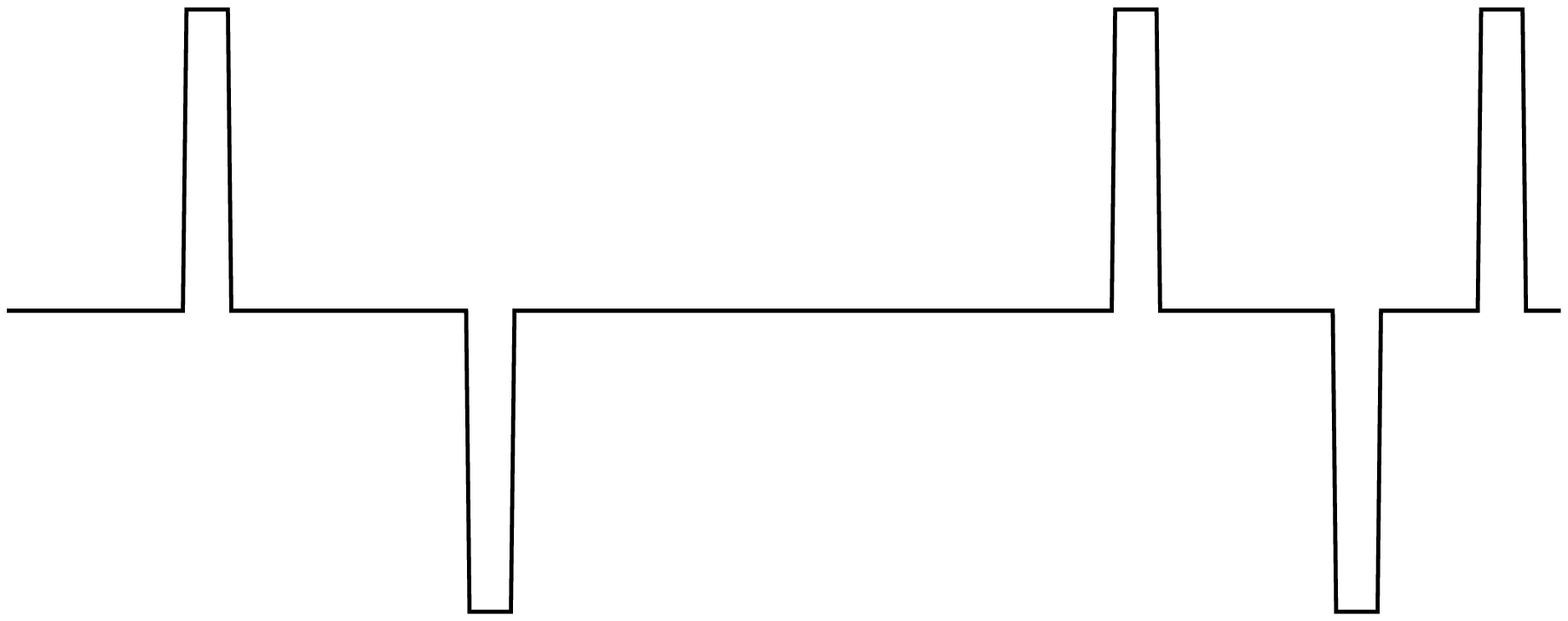}
\hspace{1 mm}
\includegraphics[width = 0.3\textwidth]{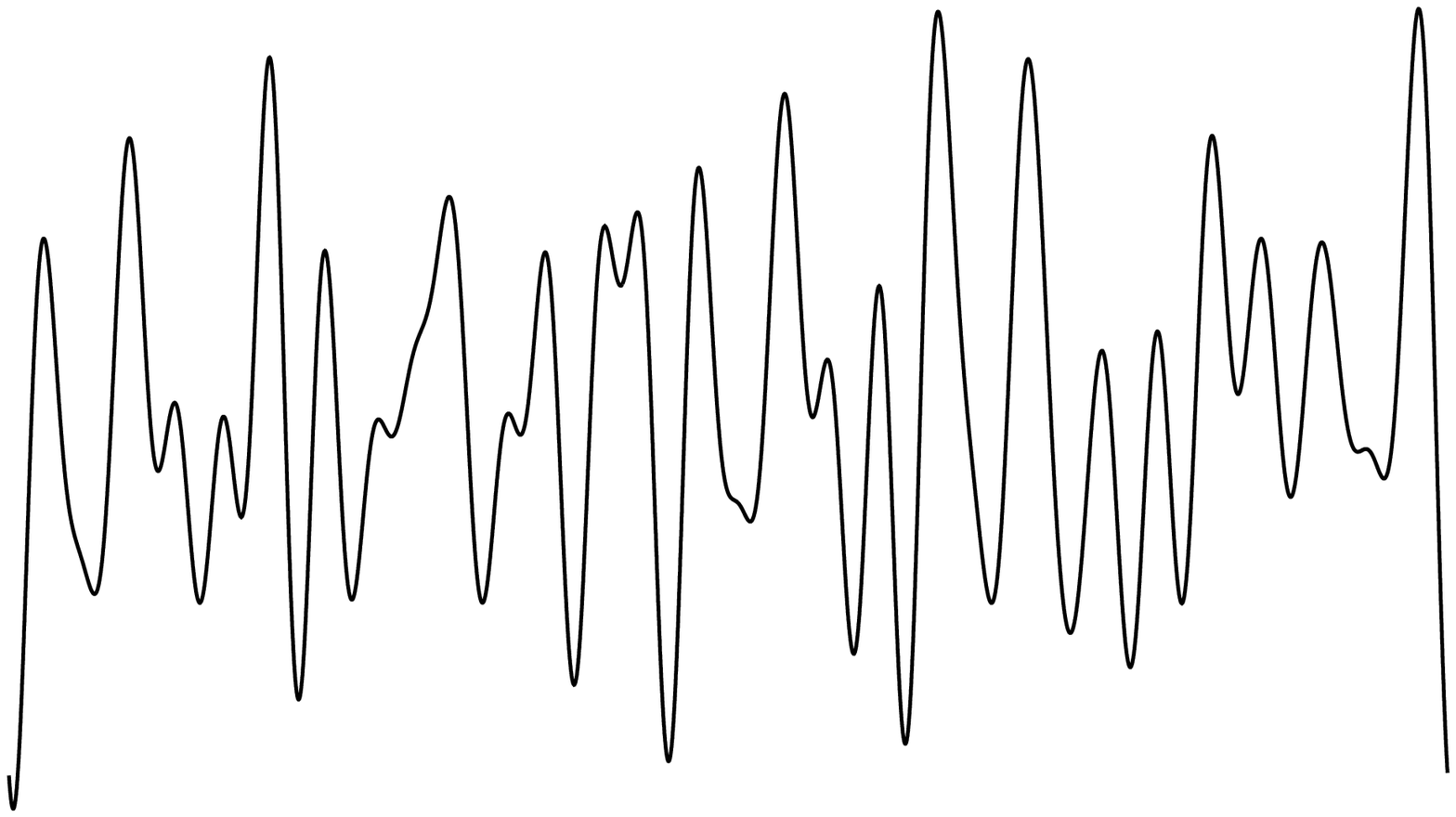}
\caption{The input signals we used: periodic pulse train, aperiodic pulse train and a band-limited noise}
\label{fig:Signals}
\end{figure}

We used three types of input signals in our simulations: the periodic pulse train for which we have already obtained high SNR gains in the double-well system \cite{Gingl01}, included here for the purposes of comparison between the wide-band and the cross-spectral gain, and two aperiodic signals, a pulse train with randomised pulse locations and a band-limited noise whose upper cut-off frequency is much smaller than the bandwidth of the noise as stochastic excitation (see Fig \ref{fig:Signals}). In the case of pulse trains, we defined the \emph{duty cycle} of the signal as $2\tau/T$, where $\tau$ is the pulse width and $T$ is the period of the periodic pulse train.

The parameters of our mixed-signal simulations are summarised in Table \ref{tab:TheParametersOfTheSimulations}. In the case of the randomised pulse train, we determined the peak locations randomly before starting the simulations and then used exactly the same waveform in each realisation during averaging, while the band-limited noise as signal was generated anew in each averaging step.
\begin{table}[!htbp]
\centering
\begin{tabular}{|l|c|c|c|}
\hline
Parameter & Periodic pulses & Aperiodic pulses & Noise as signal\\
\hline
\hline
Amplitude & \multicolumn{2}{c|}{$0.9 A_T$} & N/A\\
\hline
Pulse width & \multicolumn{2}{c|}{1.3 ms (13 data point)} & N/A\\
\hline
Duty cycle & 10\% & \multicolumn{2}{c|}{N/A}\\
\hline
Standard deviation & \multicolumn{2}{c|}{N/A} & $0.31 A_T$\\
\hline
Frequency* & 39 Hz / $4.68 \cdot 10^{-3}$ Hz & \multicolumn{2}{c|}{N/A}\\
\hline
Bandwidth* & \multicolumn{2}{c|}{N/A} & 39 Hz / $4.68 \cdot 10^{-3}$ Hz\\
\hline
Bandwidth of the additive noise* & \multicolumn{2}{c|}{5 kHz / 0.6 Hz} & variable (see graphs)\\
\hline
Sampling frequency & \multicolumn{3}{c|}{10 kHz}\\
\hline
Length of samples & \multicolumn{3}{c|}{8192}\\
\hline
Cycles per sample & \multicolumn{2}{c|}{32} & N/A\\
\hline
Averages per data sequence & \multicolumn{3}{c|}{between 10 and 50}\\
\hline
\end{tabular}
\caption{The parameters of the simulations. The frequency values marked with * are measurable on two different frequency scale (as discussed above): the first value is the analogue frequency and the second is the corresponding theoretical value. $A_T$ denotes the threshold amplitude}
\label{tab:TheParametersOfTheSimulations}
\end{table}
We also determined the threshold amplitude $A_T$ experimentally as the minimum signal amplitude at which switching between wells can occur without noise, and expressed the signal amplitude and the standard deviation of the noise as normalised by this threshold.

\section{Results}
\label{sec:Results}

\begin{figure}[!htb]
\centering
\includegraphics[width = 0.4 \textwidth]{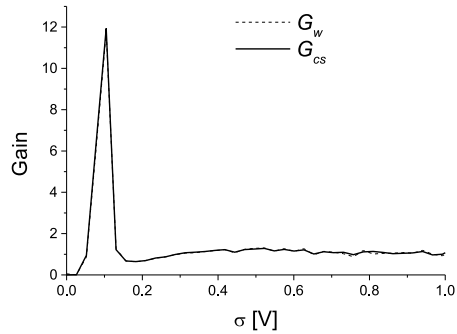}
\caption{The wide-band gain $G_w$ and the cross-spectral gain $G_{cs}$ compared in the double-well system for a periodic pulse train as input. The standard deviation of the additive input noise is denoted by $\sigma$}
\label{fig:DW-1-Periodic}
\end{figure}

First, for the purposes of validation we compared the two kinds of gains ($G_w$ and $G_{cs}$) for a periodic pulse train, in which case both are valid measures and they should theoretically yield the same results. Indeed, Fig \ref{fig:DW-1-Periodic} shows that the difference between them is negligible.

\begin{figure}[!htb]
\centering
\includegraphics[width = 0.48 \textwidth]{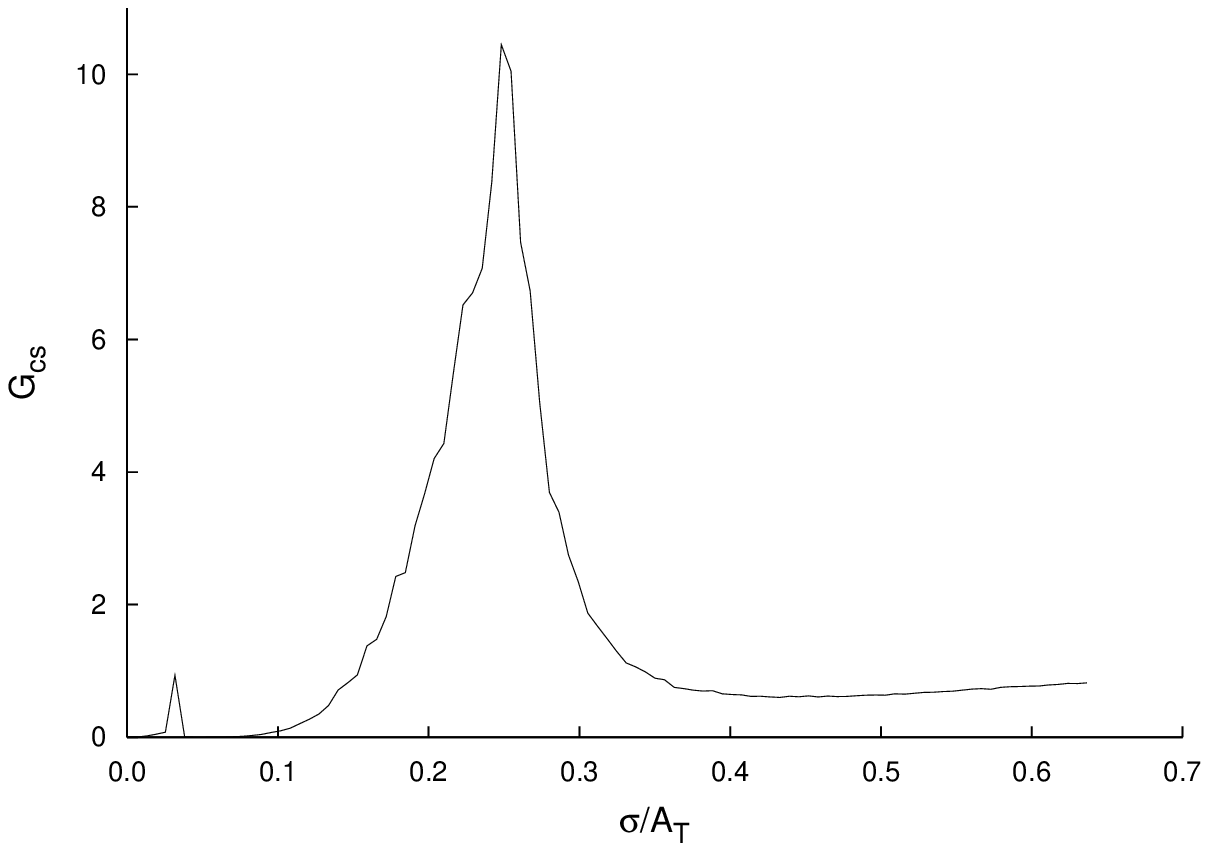}
\hspace{1 mm}
\includegraphics[width = 0.48 \textwidth]{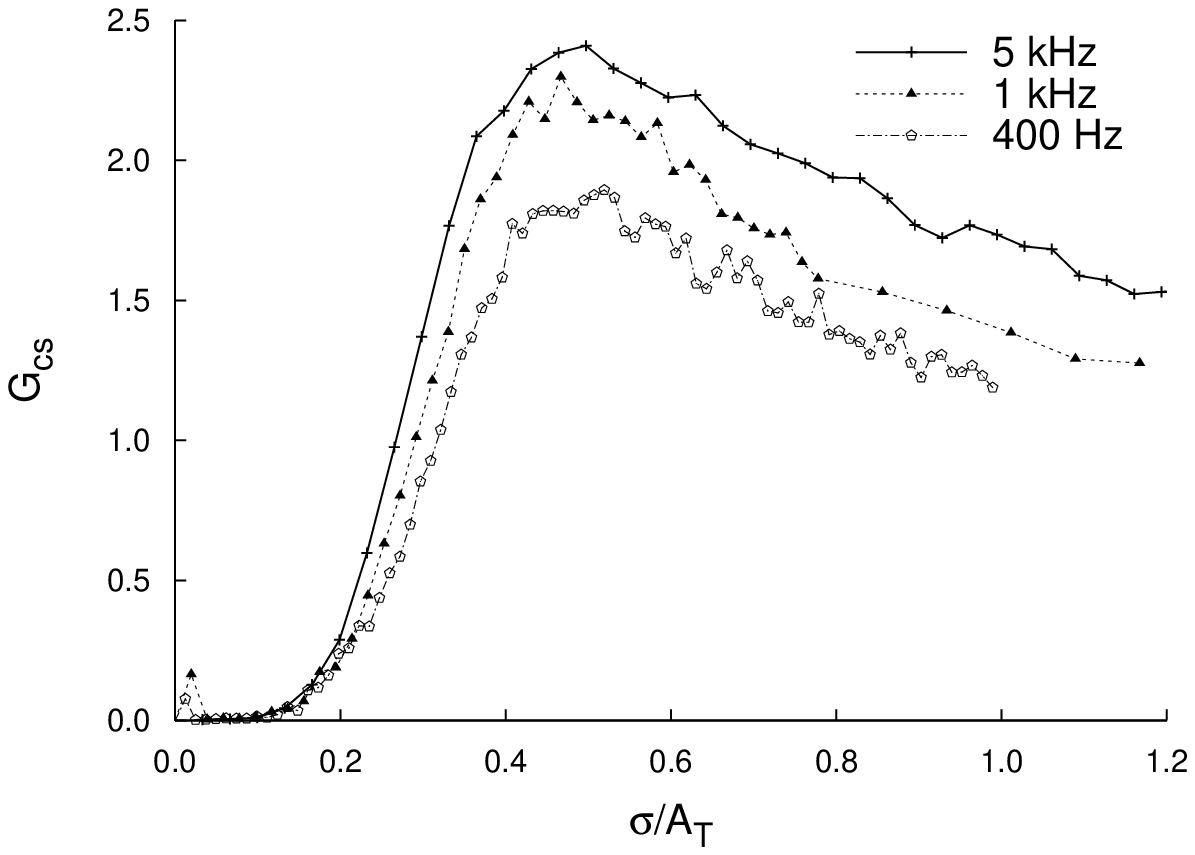}
\caption{The cross-spectral gain $G_{cs}$ in the double-well system for an aperiodic pulse train (left panel) and a band-limited noise (right panel) as input signal. The three curves on the right panel correspond to three different bandwidths of the additive noise. The standard deviation $\sigma$ of the additive input noise is normalised by the value of the threshold amplitude $A_T$}
\label{fig:DW-2-Aperiodic}
\end{figure}

The results for the aperiodic signals we were mainly interested in are depicted in Fig \ref{fig:DW-2-Aperiodic}. In the left panel, we can see that a pulse train made aperiodic by having its peaks at randomised locations can be improved by stochastic resonance almost to the same extent as its periodic counterpart. Encouraged by this finding, we went even further an applied a band-limited noise with low cut-off frequency as input signal (to avoid confusion, we use the terms \emph{noise as signal} and \emph{additive noise} to differentiate between the random process acting as input signal and the one acting as the stochastic excitation which defines stochastic resonance). From the right panel of Fig \ref{fig:DW-2-Aperiodic}, we can deduce that input-output improvement is possible even for completely random input signals.

One may argue that the input-output improvement of a noise acting as input signal stems simply from a filtering effect caused by the fact that the response time of the double well is limited, preventing high-frequency oscillations from appearing at the output. If this is the case, the major part of the additive noise gets filtered out while the noise as signal, having a much lower cut-off frequency, remains largely intact, which then leads to an input-output gain. We followed two different paths to examine this possibility: first, we compared the results obtained in the double well to those obtained in a non-dynamical stochastic resonator, the Schmitt trigger, wherein no such filtering can take place; second, we reduced the bandwidth of the additive noise to get it closer to the bandwidth of the noise as signal. The data pertaining to the Schmitt trigger come from numerical simulations carried out with the same parameters as the mixed-signal measurements.

\begin{figure}[!htb]
\centering
\includegraphics[width = 0.48 \textwidth]{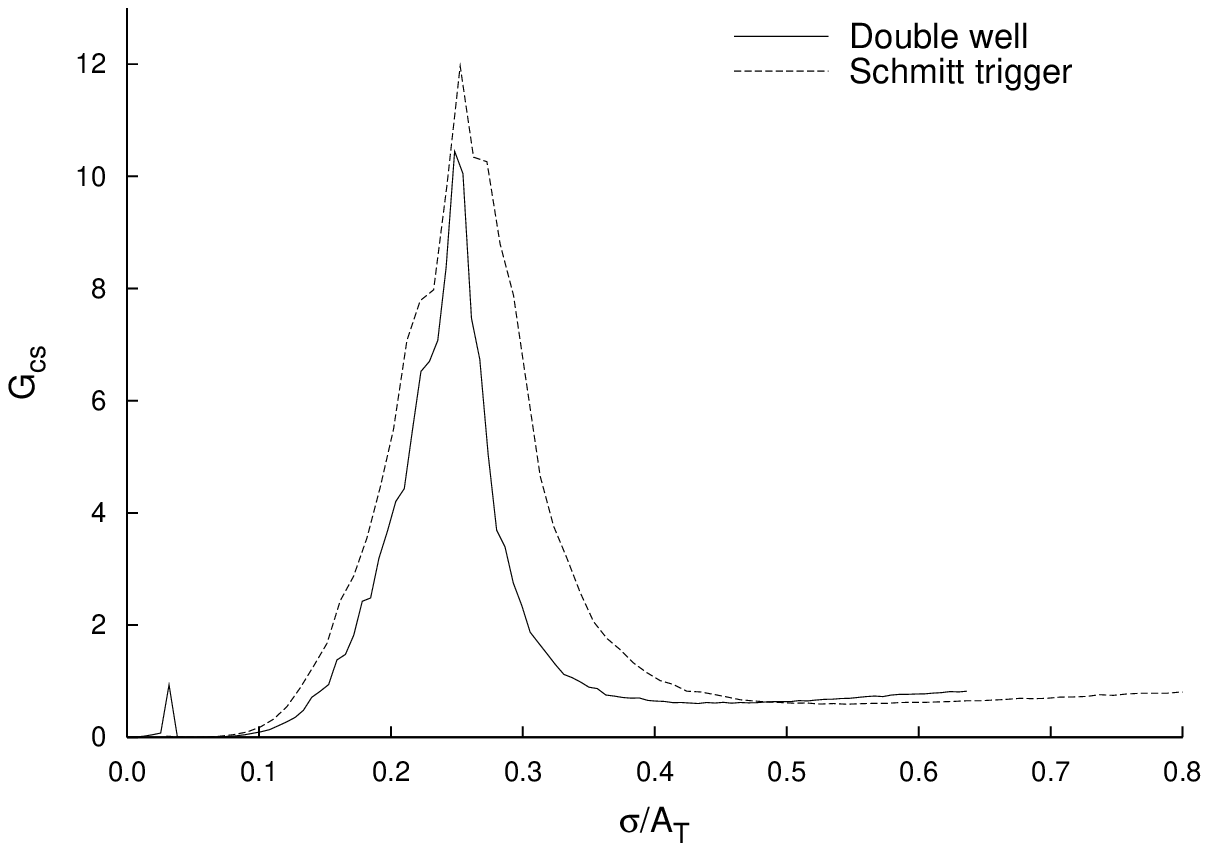}
\hspace{1 mm}
\includegraphics[width = 0.48 \textwidth]{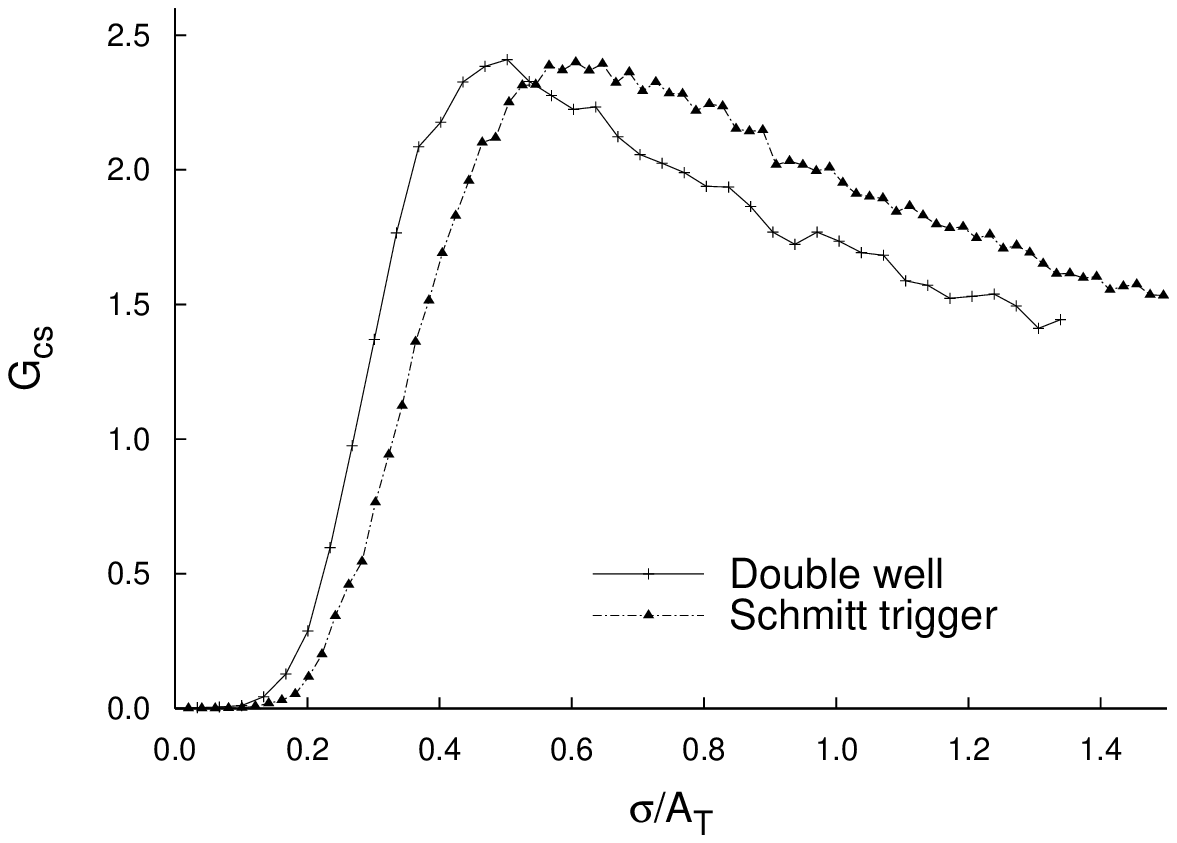}
\caption{The cross-spectral gain $G_{cs}$ in the double-well system and in the Schmitt trigger compared for an aperiodic pulse train (left panel) and a band-limited noise (right panel) as input signal. The standard deviation $\sigma$ of the additive input noise is normalised by the value of the threshold amplitude $A_T$}
\label{fig:DWST-3-Comparison}
\end{figure}

In Fig \ref{fig:DWST-3-Comparison} we can observe the similarity between the results in the double well and those in the Schmitt trigger, which suggests that the limited response time due to the dynamics of the double-well system may not play a significant role in producing an input-output gain. Indeed, at low frequencies such as those we have chosen the output of the double well is very similar to that of the Schmitt trigger.

\begin{figure}[!htb]
\centering
\includegraphics[width = 0.48 \textwidth]{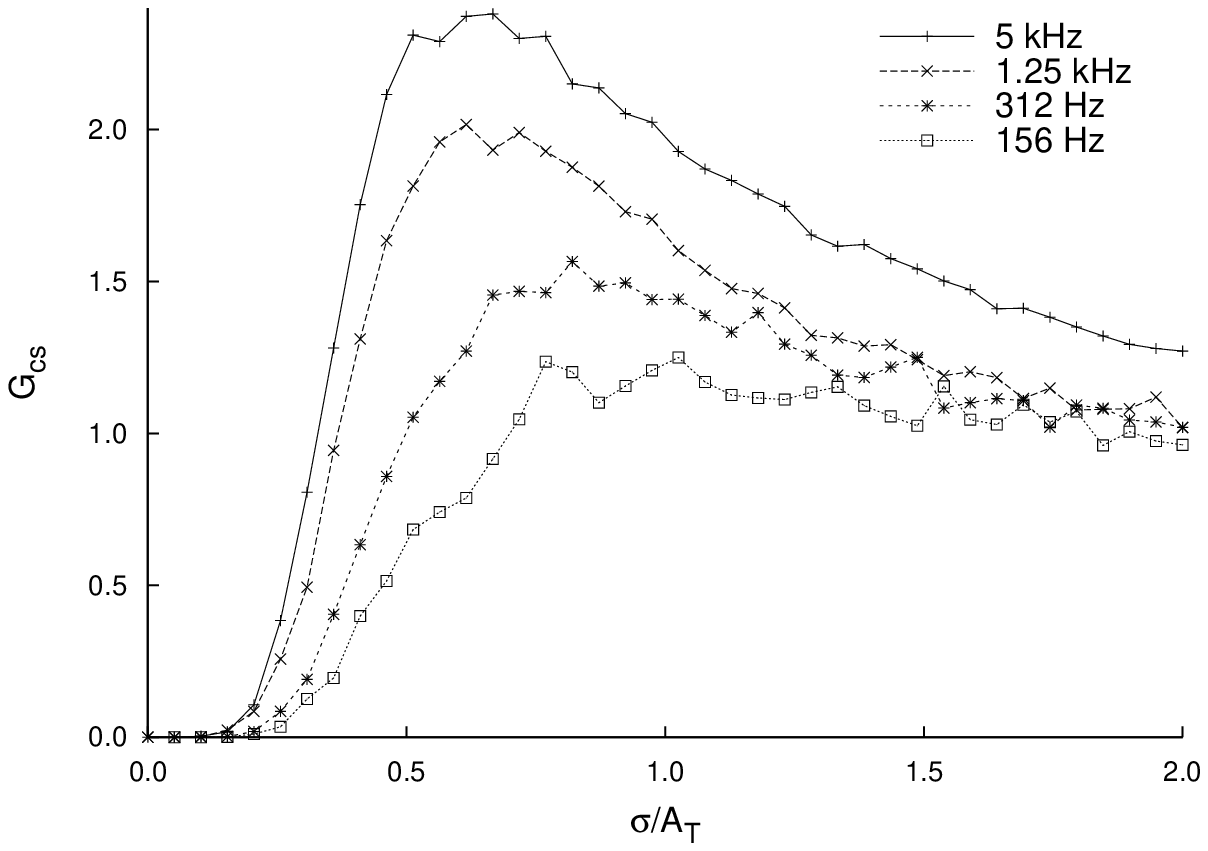}
\hspace{1 mm}
\includegraphics[width = 0.48 \textwidth]{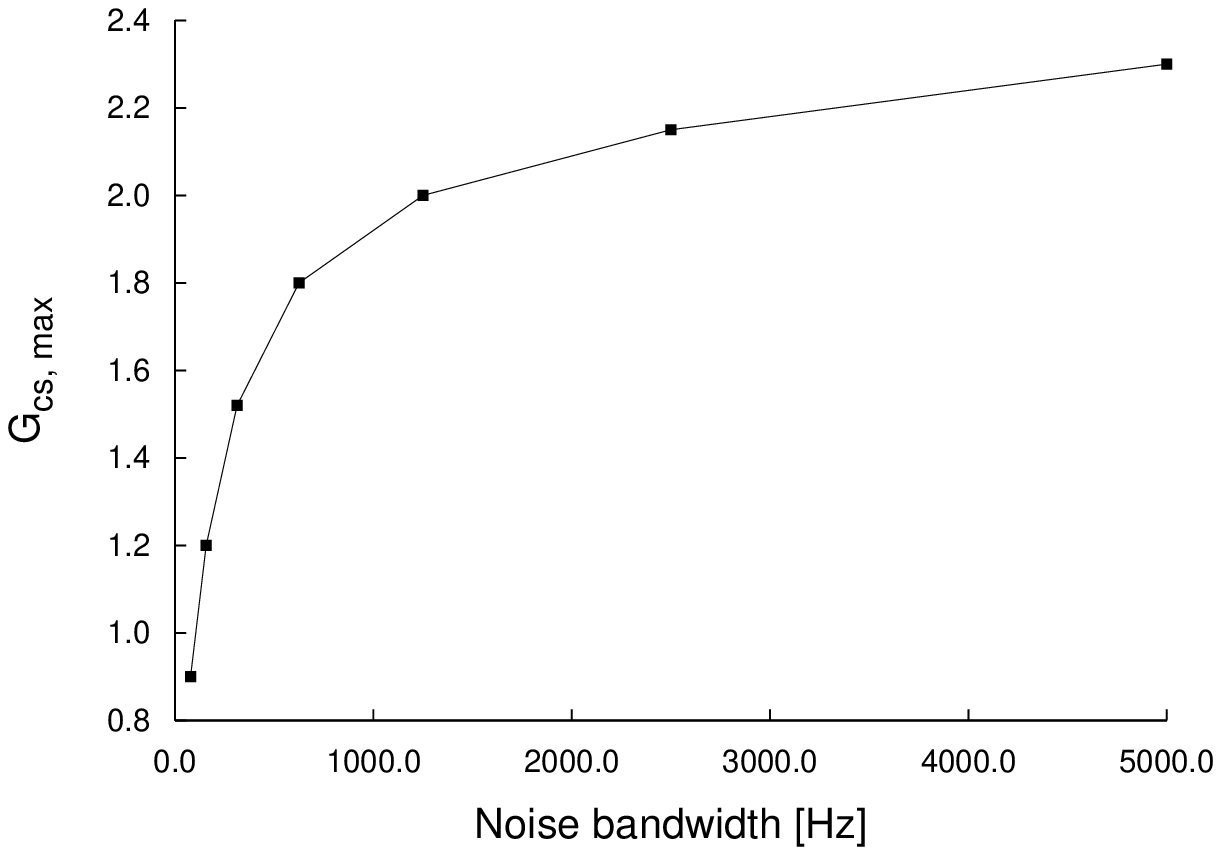}
\caption{The dependence of the cross-spectral gain $G_{cs}$ obtainable in the Schmitt trigger for a band-limited noise acting as input signal on the bandwidth of the additive noise . The right panel depicts the maximum of the gain curves as a function of the bandwidth of the additive noise}
\label{fig:ST-4-VsBandwidth}
\end{figure}

From the right panel of Fig \ref{fig:DW-2-Aperiodic} we can infer that reducing the bandwidth of the additive noise, while degrading the value of the gain, does not prevent input-output amplification itself. Comparing the right panel of Fig \ref{fig:DW-2-Aperiodic} with the left panel of Fig \ref{fig:ST-4-VsBandwidth}, we may also see that the reduction in the bandwith of the additive noise affects the two systems, the dynamical and the non-dynamical, in a very similar way, showing that the decrease in the value of the gain is not a result of a filtering effect. We also examined how the maximum of the gain depends on the bandwidth of the additive noise: the right panel of Fig \ref{fig:ST-4-VsBandwidth} shows that input-output improvement occurs in a very wide additive noise bandwidth range and the value of a gain only sinks below one for bandwidths that are less than ten times the signal bandwidth (it is important to note here that a noise bandwidth much greater than the frequency of the signal is a requirement for stochastic resonance itself to take place).

\section{Conclusions}
\label{sec:Conclusions}

Utilising a mixed-signal simulation system, we have demonstrated that the stochastic resonance occurring in the archetypal double-well model can lead to a significant input-output improvement even for aperiodic signals. We applied two kinds of aperiodic signals, a randomised pulse train and a band-limited noise as input signal, and using a cross-spectral measure to reflect their noise content both at the input and at the output, we have found input-output gains well above unity for both types of signals.

From a comparison between the dynamical double-well and the non-dynamical Schmitt trigger, and from studying the dependence of the gain on the bandwidth of the additive noise, we can conclude that the significant signal improvement we have found is not a result of a filtering effect due to the limited response time inherent in the dynamics of the double well.

Our results bring about a significant extension of the range of signal types which can be improved by stochastic resonance. Now it is clear that---although the value of the gain may depend on the type of the input signal---there are no strict requirements for the input signal to be amplified by SR: it need not be pulse-like or periodic at all; in fact, even completely random signals may be improved.

\section*{Acknowledgements}
\label{sec:Acknowledgements}

Our research has been funded by OTKA (Hungary), under grant T037664. Z Gingl acknowledges support from the Bolyai Fellowship of the Hungarian Academy of Sciences.
\bibliography{Makra}
\end{document}